%% file: proceedings.tex
\documentclass[preprint,12pt]{elsarticle}

\usepackage{amssymb}
\usepackage{amsthm}
\usepackage{amsmath}
\usepackage{axodraw2} 
\usepackage{bmpsize}
\usepackage{epsfig} 

\include{graphs}

\def\beq{\begin{equation}}   
\def\eeq{\end{equation}}
\def\bea{\begin{eqnarray}}  
\def\eea{\end{eqnarray}} 
\def\nn{\nonumber}
\def\r{\right} 
\def\l{\left} 
\def\f21{{}_2F_{1}}
\def\eps{\epsilon}

\journal{Nuclear Physics B}

\newcommand{\MSb}{$\overline{\mbox{MS}}$}
\newcommand{\ra}{\rightarrow}

\newcommand{\als}{\alpha_{\rm s}}

\def\nf{{n^{}_{\! f}}}

\def\as(#1){{\alpha_{\rm s}^{\:#1}}}
\def\ar(#1){{a_{\rm s}^{\:#1}}}

\def\eps{\epsilon}

\begin{document}

\begin{frontmatter}



\title{Zimmermann's Forest Formula, Infrared Divergences and the QCD Beta Function}


\author{Franz Herzog}
\ead{fherzog@nikhef.nl}
\address{Nikhef Theory Group,\\ Science Park 105, 1098 XG Amsterdam, The Netherlands}

\begin{abstract}
We review Zimmermann's forest formula, which solves Bogoliubov's recursive $R$-operation for the subtraction of ultraviolet divergences in perturbative Quantum Field Theory. We further discuss a generalisation of the $R$-operation which subtracts besides ultraviolet also Euclidean infrared divergences. This generalisation, which goes under the name of the $R^*$-operation, can be used efficiently to compute renormalisation constants. We will discuss several results obtained by this method with focus on the QCD beta function at five loops as well as the application to hadronic Higgs boson decay rates at N${}^4$LO. This article summarizes a talk given at the Wolfhart Zimmermann Memorial Symposium.  
\end{abstract}


\end{frontmatter}

\pagebreak

\section{Introduction}
\label{sec:intro}

Despite the enormous success in describing the interactions of elementary particles the appearance of ultraviolet (UV) divergences make it difficult to establish Quantum Field Theory as a fundamental theory of nature. A solution which at least grants the interpretation of Quantum Field Theories as low energy effective theories is given by the procedure of renormalisation. This is the procedure to absorb the troublesome infinities, present at small distances, into the physical parameters of the theory. 

An important development in the establishment of renormalisation theory has been the Bogoliubov-Parasiuk-Hepp-Zimmermann (BPHZ) renormalisation scheme. This scheme was originally developed by Bogoliubov and Parasiuk \cite{Bogoliubov:1957gp} in terms of a recursive subtraction operation, often called Bogoliubov's $R$-operation. This method of renormalisation makes it possible to subtract the complicated overlapping and nested UV-divergences which can appear in Feynman integrals, the building blocks of the perturbative expansion. 

An important aspect of the BPHZ scheme is that it connects the renormalisation constants, which are usually associated to counterterms at the Lagrangian level, to explicit local counterterms at the level of the integrands associated to Feynman graphs. Beyond this the BPHZ scheme is also practical, as it allows one (surely in massive theories and at least in principle also in massless theories) to arrive at absolutely convergent representations for renormalised Feynman integrals. Working in the BPHZ scheme Feynman integral computations can thus be carried out without the need of regulation. But also in the presence of a (dimensional or analytic) regulator the BPHZ-scheme provides an elegant way to separate the potentially complicated finite parts of Feynman integrals from their much simpler divergent parts.

Bogoliubov's and Parasiuk's proof for the finiteness of Feynman Integrals renormalised via the $R$-operation was however not completely satisfactory, and was later corrected by Hepp \cite{Hepp:1966eg}. Hepp's proof was based on a method to decompose the domain of integration of Schwinger parameters into different sectors to avoid the overlapping divergences which created difficulties in the proof of Bogoliubov and Parasiuk. 
An alternative proof for the finiteness of the renormalised Feynman Integral was given by Zimmermann 
\cite{Zimmermann:1968mu,Zimmermann:1969jj}. He realised in particular that the recursion in Bogoliubov's $R$-operation gives rise to a sum over forests of graphs. This allowed him to rewrite the $R$-operation into a form which is now often referred to as Zimmermann's forest formula. Using the combinatoric properties of forests Zimmermann formulated an elegant and comparably simple proof for the finiteness of Feyman Integrals directly in momentum space. 

This article is organised as follows. We will give an overview of further extensions of BPHZ in section \ref{sec:extension}.
In section \ref{sec:Forest} we will review the $R$-operation and in particular Zimmermann's forest formula. We will then consider an infrared (IR) generalisation of the $R$-operation in section \ref{sec:R*}. This $R^*$-operation is a powerful tool in modern computations of renormalisation constants in gauge and scalar quantum field theories. We will continue by discussing applications of this method. In section \ref{sec:beta} we will focus on the calculation of the five-loop beta function in QCD and report briefly on hadronic Higgs boson decay rates at next-to-next-to-next-to-next-to-leading order (N${}^4$LO) in perturbative QCD in section \ref{sec:Hgg}.

\section{Extensions of BPHZ}
\label{sec:extension}
The BPHZ scheme, with it's underlying forest formula, has been extended in several different directions which we attempt to briefly summarise in the following.
\subsection{BPHZL}   
One such extension, introduced by Zimmermann and Lowenstein, goes by the name of the BPHZL scheme \cite{Lowenstein:1974qt,Lowenstein:1975rg,Lowenstein:1975ps}. Since the BPHZ scheme is not properly equipped to deal with massless particles, such as gauge bosons or massless fermions, BPHZL extents BPHZ to such cases. The problem which occurs when applying the original BPHZ scheme to massless theories, is that IR-divergences could appear in the counterterms constructed by the BPHZ procedure. It is a well known fact, guaranteed by the KLN theorem \cite{Kinoshita:1962ur,Lee:1964is}, that the IR divergences are spurious and must cancel in properly defined physically measurable quantities. The BPHZL scheme carefully introduces masses into the UV counterterms in order to regulate their IR divergences, thereby avoiding the introduction of new (un-physical) IR divergences into the theory.   
\subsection{The $R^*$-operation}   
Another extension of BPHZ is the $R^*$-operation by Chetyrkin, Tkachov and Smirnov \cite{Chetyrkin:1982nn,Chetyrkin:1984xa}. The $R^*$-operation is equipped to subtract besides the UV also IR divergences of Euclidean Feynman integrals. This is how it differs substantially from the BPHZL scheme which does not intend to subtract the IR divergences, but rather tries to avoid them in UV counterterms - thereby leaving their cancellation to the mechansims beneath the KLN theorem. The $R^*$-operation should be regarded more as a mathematical trick - rather than a renormalisation scheme - which allows one to extract the renormalisation constants of Feynman integrals or correlators from maximally simple one-scale Feynman Integrals. It achieves this by making use of the technique of IR rearrangement (IRR) \cite{Vladimirov:1979zm} in dimensional regularisation. IRR essentially builds on the observation that the local counterterms in dimensional regularisation must be independent (up to polynomial dependence) of the kinematic data of Green's functions \cite{Collins:1974bg}.
\subsection{Mass divergences in Minkowski space}   
A variant of BPHZ subtraction has been extended to the subtraction of collinear divergences in hadronic collisions by van Neerven and Humpert in \cite{Humpert:1980xj}. There an explicit forest formula is given which describes how collinear divergences can be treated in an analagous diagrammatic approach to UV divergences. The subtraction of soft and collinear divergences has also been described in a BPHZ like setting by Collins in his book ``Foundations of perturbative QCD'' \cite{Collins:2011zzd}. There also exist several works by Kinoshita and his collaborators Cvitanovic and Ukawa \cite{Cvitanovic:1974sv,Kinoshita:1975ie} which describe forest formulas for the subtraction of IR divergences in the Feynman parameteric representation in $g-2$ calculations in QED.
\subsection{Hopf algebraic formulation}
A more mathematical development by Kreimer and his collaborators, notably among others Connes, is the discovery that the recursive $R$-operation gives rise to a Hopf algebra \cite{Kreimer:1997dp,Connes:1999yr}. This development has thus uncovered the fundamental  mathematical structure behind the renormalisation procedure and may shed further light into the mathematics underlying Quantum Field Theory as a whole, which is still poorly understood from a rigorous point of view. Another interesting development in this direction was made quite recently by the mathematician Brown. He showed that the Hopf algebra present in a certain linear blow-up or slicing scheme in projective Schwinger parameters can also lead to a more general Hopf algebra which covers also the infra-red divergences present in euclidean Feynman integrals \cite{Brown:2015fyf}. It remains yet to connect this work with that of the $R^*$-operation which is formulated in momentum space; although it appears clear that the two are related. For instance the notion of motic subgraphs introduced by Brown in \cite{Brown:2015fyf} is equivalent to the notion of infrared irreducibility introduced in \cite{Chetyrkin:1984xa}. In fact it has already been used to build an efficient IR-subgraph search algorithm in the context of the $R^*$-operation in momentum space in \cite{Herzog:2017bjx}. 
\section{The BPHZ $R$-operation}
\label{sec:Forest}
\subsection{Bogoliubov's recursion}
We begin with a brief review of Bogoliubov's recursive $R$-operation. We shall in the following denote by $\Gamma$ either a Feynman graph or its associated momentum space integrand or integral; thus leaving its precise meaning to be determined by the circumstances of its appearance.  

The renormalised integrand of the Feynman Graph $\Gamma$, $R(\Gamma)$, is then defined as follows:
\begin{equation}
\label{eq:R}
R(\Gamma)=\sum_{S\subseteq\Gamma} Z(S)*\Gamma/S\,,\qquad Z(S)=\prod_{\gamma\in S}Z(\gamma)\,.
\end{equation}
Here the sum goes over all spinneys $S$ contained in $\Gamma$. Spinneys are possibly disjoint sets of UV-divergent 1PI subgraphs ($\gamma$s) of $\Gamma$. A valid spinney can also be the empty graph or the full graph $\Gamma$ itself. The contracted graph $\Gamma/S$ is constructed by contracting to points in $\Gamma$ each of the disjoint 1PI components $\gamma$ in the spinney $S$. The functions $Z(\gamma)$ can be identified with the local UV counterterm - or ``renormalisation constant'' - of a 1PI subgraph $\gamma$. Here the word local signifies that $Z(\gamma)$ is a homogeneous polynomial in the external momenta and masses whose degree equals the superficial degree of divergence of the Feynman graph $\gamma$, denoted by $\omega(\gamma)$. The $*$ symbol indicates insertion of the local counterterm $Z(\gamma)$ into the vertex into which $\gamma$ was contracted in $\Gamma/S$.

The local UV-counterterm $Z(\Gamma)$ is itself defined recursively:
\begin{equation}
\label{eq:Z}
Z(\Gamma)=-K\bigg(\sum_{S\varsubsetneq\Gamma} \prod_{\gamma\in S} Z(\gamma)*\Gamma/S\,   \bigg)\,,
\end{equation}
where the sum goes over all spinneys $S$ which do not contain the full graph $\Gamma$.
The $K$-operation is defined to isolate, according to a certain renormalisation scheme, the singular part of its argument. For example the $K$-operation, in the minimal subtraction (MS) scheme, acting on a meromorphic function $F(\epsilon)$ will return only the pole part of $F(\epsilon)$. In a momentum subtraction scheme the $K$-operation may instead act as a Taylor expansion operator around a certain fixed or vanishing external momentum configuration. Thus, while the general form of the $R$-operation is unique, the precise value of the local counterterm $Z$ is scheme dependent. Let us further remark that for the empty subgraph, which we will denote by $\emptyset$, one requires $Z(\emptyset)=1$.

\begin{figure}  
\centering
\includegraphics[width=1 \textwidth,clip,trim= 0 12cm 0 2cm]{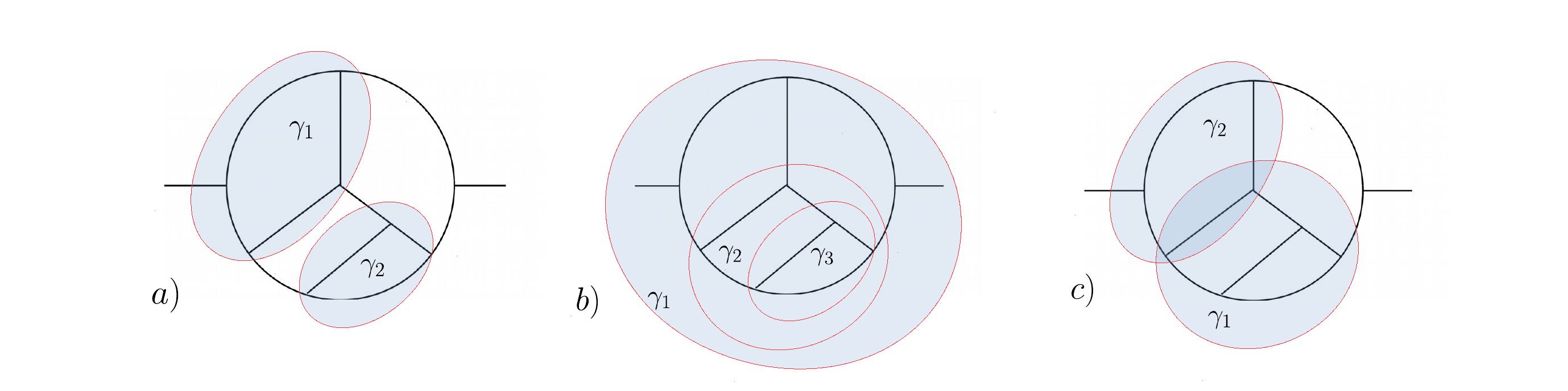}
\caption{This figure shows several examples of 1PI subgraphs. $a)$ shows an example of a $\Gamma$-forest which is also a spinney, since the two subgraphs $\gamma_1$ and $\gamma_2$ are disjoint. $b)$ shows an example of a $\Gamma$-forest which is not a spinney since $\gamma_3\subset\gamma_2\subset\gamma_1$. The two subgraphs $\gamma_1$ and $\gamma_2$ in $c)$ are neither a spinney nor a $\Gamma$-forest since they overlap.}
\label{fig:forests}
\end{figure}

\subsection{Zimmermann's Forest Formula}
Zimmermann's forest formula arises as a consequence of the recursive structure of the $R$-operation.
The recursion essentially generates nested sets of subgraphs. The nested structures thus arising are called $\Gamma$-forests by Zimmermann. A $\Gamma$-forest $U(\Gamma)$ is a set of subgraphs of $\Gamma$, which are either disjoint or are subgraphs of one another, or more abstractly
\begin{equation}
U(\Gamma)=\{\gamma_1,...,\gamma_n\,|\gamma_i\in \Gamma\;\text{and}\;(\gamma_i\cap\gamma_j=\emptyset\;\text{or}\;\gamma_i\subset\gamma_j\;\text{or}\;\gamma_i\supset\gamma_j)\}\,.
\end{equation}
If a $\Gamma$-forest contains only UV divergent subgraphs it is called a restricted $\Gamma$-forest
and is denoted by $U_r(\Gamma)$. Let us remark at this point that with this definition a spinney is always a restricted $\Gamma$-forest but a restricted $\Gamma$-forest is not always a spinney. Some examples for $\Gamma$-forests and spinneys are illustrated in figure \ref{fig:forests}.

Let us further define the set $\mathcal{U}_r(\Gamma)$ as the set of all restricted $\Gamma$-forests. The $R$-operation can then be written as Zimmermann's forest formula:
\begin{equation}
R(\Gamma) = \sum_{U\in \mathcal{U}_r(\Gamma)}\,\prod_{\gamma\in U}\,(-K_\gamma)\;\Gamma \,, 
\end{equation}
where $K_\gamma\,\Gamma=K(\gamma)*\Gamma/\gamma$ and if the forest contains nested subgraphs the operations $K_\gamma$ should be applied from inside out.
\subsection{Example}
Let us consider the following Feynman integral as an example:
\begin{eqnarray}
&&\Gamma=\bubbletwok\\
&&=\int \frac{d^Dk_1}{i\pi^{D/2}}\frac{d^Dk_2}{i\pi^{D/2}}\frac{1}{(k_1^2+m^2)((k_1+k_2)^2+m^2)(k_2^2+m^2)((k_1+p)^2+m^2)} \nn  
\end{eqnarray}
Acting with the $R$-operation in a momentum subtraction scheme, where the action $K$ onto a graph corresponds to a Taylor expansion around vanishing external momenta, yields:
\begin{eqnarray}
R \l(\bubtwonum{}{}{}{} \r)&=& \bubtwonum{}{}{}{} -K\l(\bubtwonum{}{}{}{}\r)- K\l(\bubbleonenum{}{}\r) \cdot \bubbleonenum{}{}\nn\\
&&+K\l(K\l(\bubbleonenum{}{}\r)\cdot\bubbleonenum{}{}\r)\\
&=& \bubtwonum{}{}{}{} -\vactwonum{}{}{}{}- \vaconenum{} \cdot \bubbleonenum{}{}+\vaconenum{}\cdot\vaconenum{}\nn
\end{eqnarray}

\section{The $R^*$-operation}
\label{sec:R*}
The $R^*$-operation acting on a Euclidean Feynman graph $\Gamma$ can be written as 
\begin{equation}
R^* (\Gamma) = \sum_{\substack{S \subseteq \Gamma,\tilde S \subseteq \Gamma \\S \cap \tilde S = \emptyset}}\;\widetilde Z(\tilde S)* Z(S) * \Gamma/S \setminus \tilde S\,.
\end{equation}
Here the sum goes over disjoint pairs of UV and IR spinneys $S$ and $\tilde S$ respectively. The UV spinney is defined identically as in the case of the $R$-operation. To define the IR spinney $\tilde S$ is slightly more involved than for UV spinneys and is for this reason referred to the literature \cite{Chetyrkin:1984xa,Chetyrkin:2017ppe,Herzog:2017bjx}. The remaining contracted graph $\Gamma/S\setminus\tilde S$ is constructed by first contracting the $S$ in $\Gamma$ and then deleting the lines and vertices contained in $\tilde S$ in $\Gamma/S$. The case in which $\tilde S=\Gamma$ can occur only if $\Gamma$ is a scaleless vacuum graph of logarithmic superficial degree of divergence. In this case $\Gamma\setminus \tilde S$ is defined as the unit $1$. The UV and IR counterterm operations $Z$ and $\tilde Z$ are then defined recursively via:
\beq
Z (\Gamma) = -K\Big(\sum_{\substack{S \subsetneq \Gamma,\tilde S \subseteq \Gamma \\S \cap \tilde S = \emptyset}}\;\widetilde Z(\tilde S)* Z(S) * \Gamma/S \setminus \tilde S\Big)\,,
\eeq
where one omits in the sum over UV spinneys the full graph $\Gamma$ and
\beq
\tilde Z (\Gamma_0) = -K\Big(\sum_{\substack{S \subseteq \Gamma_0,\tilde S \subsetneq \Gamma_0 \\S \cap \tilde S = \emptyset}}\;\widetilde Z(\tilde S)* Z(S) * \Gamma_0/S \setminus \tilde S\Big)\,,
\eeq
where one omits in the sum over IR spinneys the scaleless vacuum Feynman graph $\Gamma_0$. The identity $R^*(\Gamma_0)=0$ can be used to find relations among IR and UV counterterms in dimensional regularisation.
\subsection{Example}
Let us consider the Feynman integral:
\beq
\Gamma=\bubtwodotnum{1}{2}{3}=\int \frac{d^Dk_1}{i\pi^{D/2}}\frac{d^Dk_2}{i\pi^{D/2}}\frac{1}{(k_1^2)^2(k_2+P)^2(k_1+k_2)^2}
\eeq
Here we have labeled the lines from 1-3, such that their corresponding momenta are parameterised as 
$q_1=k_1,q_2=k_2+P,q_3=k_1+k_2$ respectively. The example features an IR divergence when the momentum is flowing through the 
dotted line $1$ vanishes. It also features two UV divergent subgraphs, corresponding to the full graph or the subgraph which consists of lines $2$ and $3$. The action of the $R^*$ operation yields: 
\begin{eqnarray}
R^*\l(\bubtwodotnum{1}{2}{3}\r)&=&\bubtwodotnum{1}{2}{3}
+Z\l( \bubtwodotnum{1}{2}{3}\r)
+Z\l( \bubbleonenum{2}{3}\r)* \tadonenum{1}\\
&&+\tilde Z\l(\IRonenum{1}\r)*\bubbleonenum{2}{3}
+\tilde Z\l(\IRonenum{1}\r)*Z\l( \bubbleonenum{2}{3}\r)*1\,.\nn
\end{eqnarray}
The IR counterterm can be evaluated as 
\beq
\tilde Z\l(\IRonenum{}\r)=\tilde Z\l(\vaconenum{}{}\r) = -Z\l(\vaconenum{}{}\r)=K\l(\bubbleonenum{}{} \r)\,.
\eeq

\section{The QCD beta function at five loops}
\label{sec:beta}
\begin{figure}  
\centering
\includegraphics[width=1 \textwidth,clip,trim= 0 14cm 0 2cm]{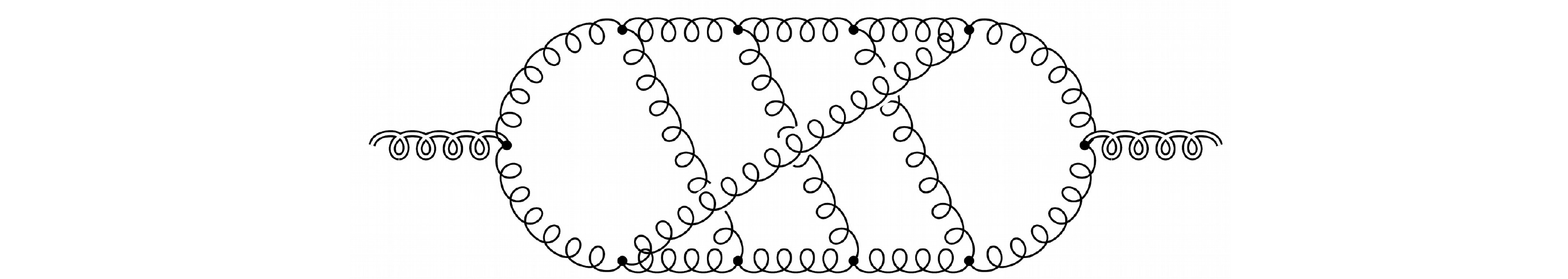}
\caption{This figure shows a typical five-loop Feynman graph contributing to the background field self energy.}
\label{fig:Benergy}
\end{figure}

\begin{figure}  
\centering
\includegraphics[width=1 \textwidth,clip,trim= 0 14cm 0 2cm]{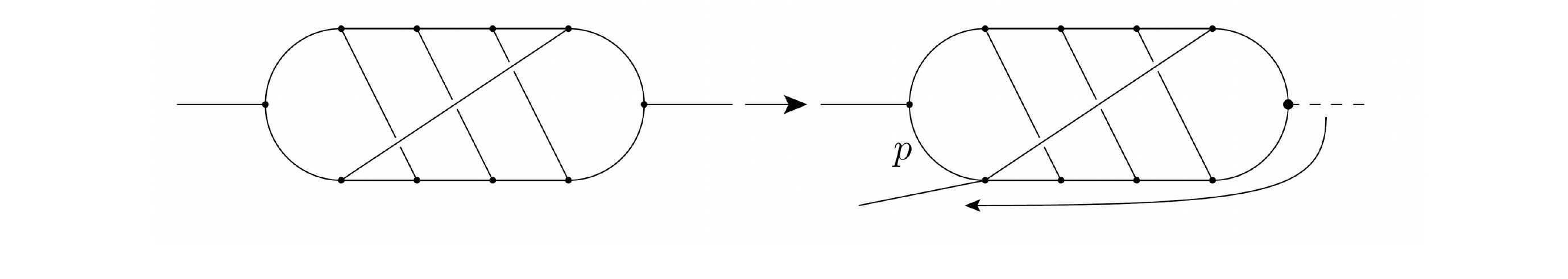}
\caption{This figure illustrates the procedure of IR rearrangement to simplify the calculation of UV poles of Feynman integrals.}
\label{fig:IRR}
\end{figure}

By governing the scale evolution of the (reduced) strong coupling constant,
\beq
a(\mu)=\frac{\alpha_s(\mu)}{4\pi}=\l(\frac{g_s(\mu)}{4\pi}\r)^2 \,,
\eeq
the beta function (in the MS scheme of dimensional regularisation \cite{DimReg2}),
\begin{equation}
\beta(a)  = \frac{d a}{d\log \mu^2}=-\sum_{n=0}^\infty \beta_n a^{n+2}\,, 
\end{equation}
is of fundamental importance to QCD. Pioneering calculations of the 1-loop beta function in the 60s and early 70s \cite{AF1,AF2,beta0tH,beta0a,beta0b} lead to the discovery of asymptotic freedom in QCD. Since then tremendous progress in perturbative calculations has lead to the determination of the beta function up to five loops \cite{beta1a,beta1b,beta1c,  beta2a, beta2b, beta3a, beta3b, beta4SU3, beta4nf3, Herzog:2017ohr, Chetyrkin:2017bjc, Gracey:1996he, Luthe:2017ttg}. Several of these calculations were based on the use of the $R^*$-operation which we briefly discussed in section \ref{sec:R*}. Here we wish to report in particular on the calculation of ref~\cite{Herzog:2017ohr}. This calculation made use of the background field method, where the renormalisation constant, associated to the background field $Z_B$, is related to that of the running coupling via \cite{Abbott80,AbbottGS83}:
\beq
Z_aZ_B=1
\eeq
This relation allows the extraction of $Z_a$ and therefore also $\beta$ at five loops from the knowldege of the poles of the 5-loop background field self energy (see figure \ref{fig:Benergy}). 
Utilising the techniques of IR rearrangement (see figure \ref{fig:IRR}) and the $R^*$ methods introduced in \cite{Herzog:2017bjx} we were able to extract these poles from five-loop Feynman graphs which can be factorised into products of trivial one-loop graphs times four-loop graphs, which in turn can be computed efficiently with the {\sc Forcer} program \cite{tuACAT2016,tuLL2016,FORCER}. In order to perform these calculations we build a computational framework based on {\sc FORM} \cite{FORM3,TFORM,FORM4}, QGRAF \cite{QGRAF} for the generation of diagrams and the colour package of ref~\cite{Colour}.

Let us now present some numeric results for the beta-function for the gauge group SU($3$), fixing also the number of quark flavours $n_f$ to a few physically relevant values:
\bea
\label{bSnf3}
  \widetilde{\beta}(\als,\nf\!=\!3) &\! =\! &
       1
       + 0.565884 \,\* \als
       + 0.453014 \,\* \as(2)
       + 0.676967 \,\* \as(3)
       + 0.580928 \,\* \as(4)
\; , \quad \nn \\[1mm]
  \widetilde{\beta}(\als,\nf\!=\!4) &\! =\! &
       1
       + 0.490197 \,\* \als
       + 0.308790 \,\* \as(2)
       + 0.485901 \,\* \as(3)
       + 0.280601 \,\* \as(4)
\; , \quad \nn \\[1mm]
  \widetilde{\beta}(\als,\nf\!=\!5) &\! =\! &
       1
       + 0.401347 \,\* \als
       + 0.149427 \,\* \as(2)
       + 0.317223 \,\* \as(3)
       + 0.080921 \,\* \as(4)
\; , \quad \nn \\[1mm]
  \widetilde{\beta}(\als,\nf\!=\!6) &\! =\! &
       1
       + 0.295573 \,\* \als
       - 0.029401 \,\* \as(2)
       + 0.177980 \,\* \as(3)
       + 0.001555 \,\* \as(4)
\; , \quad \nn \\[1mm]
\eea
where $\widetilde{\beta} \equiv - \beta(a_{\rm s}) / (\ar(2) \beta_0)$. These numbers indeed show excellent perturbative convergence of the beta-function across this physically relevant range of $n_f$. It appears in particular that the convergence is enhanced for increasing values of $n_f$ at the five-loop level. It would indeed be very interesting to see whether such a pattern would continue at yet higher orders or represents a mere accident at the five loop level.

\begin{figure}  
\centering
\includegraphics[width=0.8 \textwidth,clip,trim= 0 0cm 0 0cm]{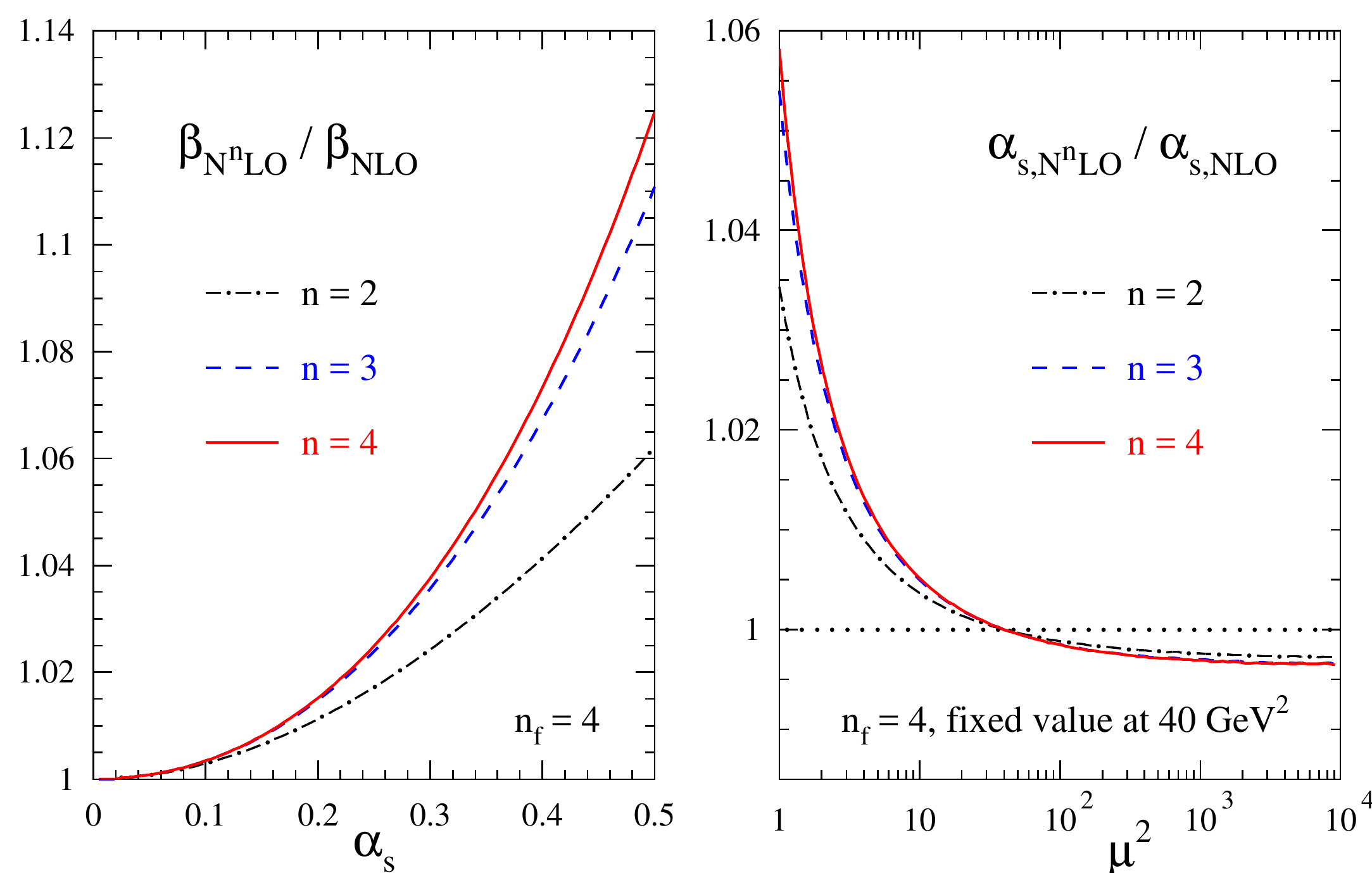}
\caption{Left panel: The total 3-,4- and 5--loop results results for the beta 
 function of QCD for four flavours, normalized to the 2-loop approximation.
 Right panel: The resulting scale dependence of $\als$ for a value of 0.2 at 
 $40 \mbox{ GeV}^2$, also normalized to the 2-loop result in order to show the 
 small higher-order effects more clearly, for the scale range 
 $1 \mbox{ GeV}^2 \leq \mu^2 \leq 10^{\,4} \mbox{ GeV}^2$.}
\label{fig:betalp}
\end{figure}

In figure \ref{fig:betalp} we show two plots illustrating the small effect of the five-loop coefficient on the scale evolution of the coupling. These effects are indeed rather mild, and show that the coupling evolution even at lower scales (which correspond to larger values of the coupling) appears now to be under excellent control. 

\section{Hadronic Higgs decays and the R-ratio at N${}^4$LO}
\label{sec:Hgg}
\begin{figure}  
\centering
\includegraphics[width=0.8 \textwidth,clip,trim= 0 0cm 0 0cm]{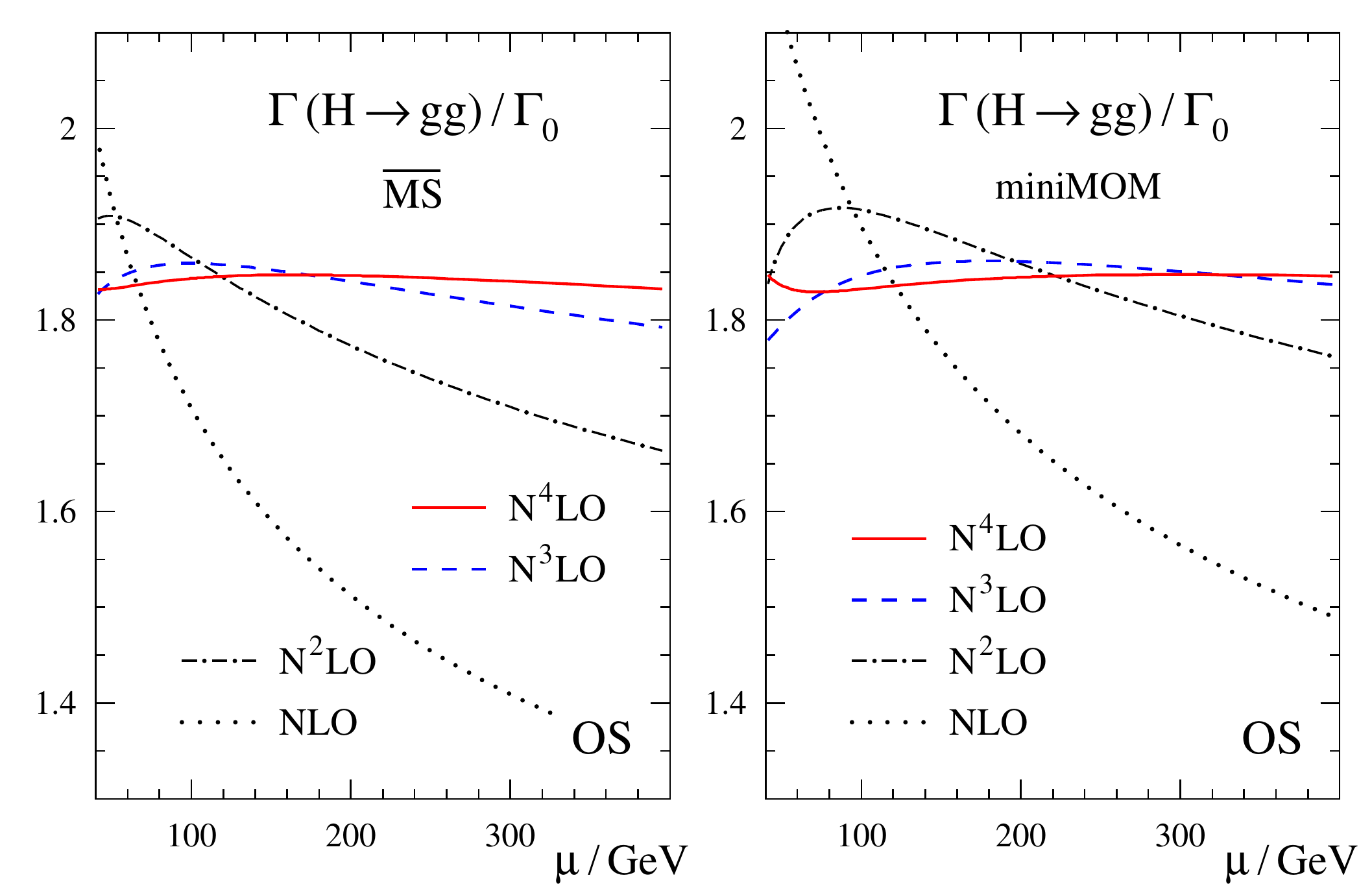}
\caption{The renormalization scale dependence of the decay width 
 $\Gamma_{H \ra\,gg}$, for an on-shell top mass of 173 GeV in \MSb\ 
 and the miniMOM scheme.}
\label{fig:hgg}
\end{figure}

Let us briefly report here also on several N${}^4$LO calculations which were completed in ref~\cite{Herzog:2017dtz} using 
essentially the same techniques which we described for the calculation of the five-loop beta function  
in section \ref{sec:beta}. Namely these are the calculations of the Higgs boson decay rate into massless bottom quarks, the Higgs boson decay rate into gluons in the heavy top quark effective theory and the hadronic $R$-ratio mediated by an off-shell photon, all of which in massless QCD. These calculations were made possible by the use of unitarity which allows one to relate decay rates to the imaginary part of a corresponding self energy. Since such self energy diagrams are real valued in Euclidean space, their imaginary pieces are suppressed by a factor of $\eps$ which originates from the analytic continuation of the phase 
\beq
\mathrm{Im} (-p^2-i\delta)^{-L\eps}=L\pi\eps\l(1-\frac{(L\pi\eps)^2}{3!}+...  \r)(p^2)^{-L\eps}\,.
\eeq
This suppression factor $(L\pi\eps)$ allows one to extract the decay rate from the UV divergences of the self energy to which it is related by unitarity. Consequentially the $R^*$-method can be applied to this problem. Since the result for the gluonic Higgs boson decay rate is new let us present here the variation of the renormalisation scale to emphasise its perturbative convergence in two different renormalisation schemes in figure \ref{fig:hgg}. Although different patterns are observed at lower orders both schemes converge to the same numerical values at this high perturbative order providing further confidence for the reliability of perturbation theory in QCD.
\section{Summary}
In this talk we reviewed the basic concepts behind Zimmermann's forest formula 
and emphasised its various extensions. In particular we discussed the $R^*$-operation, 
which generalises the forest formula to the subtraction of Euclidean infrared divergences, 
and its applications in multi-loop calculations of anomalous dimensions and decay rates.
We also discussed results for the five-loop beta function and briefly mentioned results for 
hadronic Higgs boson decay rates which were obtained using this technique. 




\section*{Acknowledgements}
This work has been supported by the {\it European Research Council}$\,$ (ERC) Advanced Grant 320651, {\it HEPGAME}.

\end{document}

%% file: graphs.tex
\def\bubbleonenum#1#2{
\raisebox{-7pt}
{
\begin{axopicture}{(30,20)(-13,-10)}
\SetScale{1}\SetColor{Blue}%
\Line(-15,0)(-10,0)
\Line(10,0)(15,0) 
\CArc(0,0)(10,0,360)
\Vertex(-10,0){1.5}
\Vertex(10,0){1.5}
\SetScale{1}\SetColor{Black}%
\Text(1.5,6.5){\tiny $#1$ \tiny}
\Text(1.5,-6.5){\tiny $#2$ \tiny}
\end{axopicture}
}
}

\def\bubbletwok{
\raisebox{-21pt}
{
\begin{axopicture}{(70,50)(-30,-24)}
\SetScale{2}\SetColor{Blue}%
\Line[arrow,arrowscale=0.5](-15,0)(-10,0)
\Line[arrow,arrowscale=0.5](10,0)(15,0) 
\CArc[arrow,arrowscale=0.5](0,0)(10,0,90)
\CArc[arrow,arrowscale=0.5](0,0)(10,90,180)
\CArc[arrow,arrowscale=0.5](0,0)(10,180,360)
\CArc[arrow,arrowscale=0.5](10,10)(10,180,270)
\Vertex(-10,0){1.0}
\Vertex(10,0){1.0}
\Vertex(0,10){1.0}
\SetScale{2}\SetColor{Black}%
\Text(-5,5){\tiny $k_1$ \tiny}
\Text(2,2){\tiny $k_2$ \tiny}
\Text(14,8){\tiny $k_1+k_2$ \tiny}
\Text(1,-7){\tiny $k_1+p$ \tiny}

\end{axopicture}
}
}

\def\bubtwonum#1#2#3#4{
\raisebox{-7pt}
{
\begin{axopicture}{(30,20)(-13,-10)}
\SetScale{1}\SetColor{Blue}%
\Line(-15,0)(-10,0)
\Line(10,0)(15,0) 
\CArc(0,0)(10,0,360)
\CArc(10,10)(10,180,270)
\Vertex(-10,0){1.5}
\Vertex(10,0){1.5}
\Vertex(0,10){1.5}
\SetScale{1}\SetColor{Black}%
\Text(-5,4){\tiny $#1$ \tiny}
\Text(2,1){\tiny $#2$ \tiny}
\Text(6,6){\tiny $#3$ \tiny}
\Text(0,-7){\tiny $#4$ \tiny}
\end{axopicture}
}
}

\def\bubtwodotnum#1#2#3{
\raisebox{-7pt}
{
\begin{axopicture}{(30,20)(-13,-10)}
\SetScale{1}\SetColor{Blue}%
\Line(-15,0)(-10,0)
\Line(10,0)(15,0) 
\CArc(0,0)(10,0,360)
\Line(-10,0)(10,0)
\Vertex(-10,0){1.5}
\Vertex(10,0){1.5}
\Vertex(0,10){1.5}
\SetScale{1}\SetColor{Black}%
\Text(1,14){\tiny $#1$ \tiny}
\Text(1,3){\tiny $#2$ \tiny}
\Text(1,-7){\tiny $#3$ \tiny}
\end{axopicture}
}
}

\def\IRonenum#1{
\raisebox{-7pt}
{
\begin{axopicture}{(10,20)(-4,-10)}
\SetScale{1}\SetColor{RedViolet}%
\Line(0,-10)(0,10)
\CCirc(0,-10){1.5}{RedViolet}{White}
\Vertex(0,0){1.5}
\CCirc(0,10){1.5}{RedViolet}{White}
\SetScale{1}\SetColor{Black}%
\Text(6,0){\tiny $#1$ \tiny}
\end{axopicture}
}
}

\def\vaconenum#1{
\raisebox{-7pt}
{
\begin{axopicture}{(30,20)(-13,-10)}
\SetScale{1}\SetColor{Blue}%
\CArc(0,0)(10,0,360)
\Vertex(-10,0){1.5}
\Vertex(10,0){1.5}
\SetScale{1}\SetColor{Black}%
\Text(1.5,-6.5){\tiny $#1$ \tiny}
\end{axopicture}
}
}

\def\vactwonum#1#2#3#4{
\raisebox{-7pt}
{
\begin{axopicture}{(30,20)(-13,-10)}
\SetScale{1}\SetColor{Blue}%
\CArc(0,0)(10,0,360)
\CArc(10,10)(10,180,270)
\Vertex(-10,0){1.5}
\Vertex(10,0){1.5}
\Vertex(0,10){1.5}
\SetScale{2}\SetColor{Black}%
\Text(-5,5){\tiny $#1$ \tiny}
\Text(2,2){\tiny $#2$ \tiny}
\Text(6,6){\tiny $#3$ \tiny}
\Text(0,-8){\tiny $#4$ \tiny}
\end{axopicture}
}
}

\def\tadonenum#1{
\raisebox{-7pt}
{
\begin{axopicture}{(30,20)(-13,-10)}
\SetScale{1}\SetColor{Blue}%
\CArc(0,5)(5,0,360)
\Line(-10,0)(10,0)
\Vertex(0,0){1.5}
\Vertex(0,10){1.5}
\SetScale{1}\SetColor{Black}%
\Text(1.5,5){\tiny $#1$ \tiny}
\end{axopicture}
}
}